\documentclass[10pt,showpacs,twocolumn,nofootinbib,aps,prd]{revtex4-1}
\usepackage{amsmath,amssymb,bm,mathrsfs}
\usepackage{mathtools,mathptmx,array,slashed}
\usepackage{graphicx,graphics,subfigure}
\usepackage{ctable,multirow}
\usepackage{tabularx}
\usepackage[colorlinks=true,linkcolor=blue,citecolor=blue,urlcolor=blue]{hyperref}
\usepackage{color}

\DeclareSymbolFont{symbols} {OMS}{cmsy}{m}{n}

\def\be{\begin{equation}}
\def\ee{\end{equation}}
\def\bea{\begin{eqnarray}}
\def\eea{\end{eqnarray}}
\def\ba{\begin{aligned}}
\def\ea{\end{aligned}}
\def\nn{\nonumber}
\def\p{\partial}

\begin{document}

%\begin{CJK*}{GBK}{song}

\title{$\widetilde{W}^{1+}$ subclass: Extending the topological classification of black hole thermodynamics}

\author{Wangyu Ai}

\author{Di Wu}
\email{Contact author: wdcwnu@163.com}
%https://orcid.org/0000-0002-2509-6729

\affiliation{School of Physics and Astronomy, China West Normal University, Nanchong, Sichuan 637002, People's Republic of China}

\date{\today}

\begin{abstract}
In this paper, we identify a novel topological subclass, dubbed $\widetilde{W}^{1+}$, in the thermodynamics of higher odd-dimensional, multiply rotating Kerr-AdS black holes. This discovery extends the established topological classification beyond the five classes and two subclasses previously known. The $\widetilde{W}^{1+}$ subclass exhibits a unique and previously unreported stability profile: it admits a thermodynamically stable small black hole state in the low-temperature limit, while in the high-temperature limit, the phase space simultaneously contains one stable large black hole, one stable small black hole, and one unstable small black hole state. Our analysis, which treats black hole solutions as topological defects, reveals a richer landscape of black hole thermodynamics than previously understood and necessitates an expansion of the topological classification scheme to accommodate this new phenomenology.
\end{abstract}

\maketitle

%\end{CJK*}

%%%%%%%%%%%%%%%%%%%%%%
\section{Introduction}
%%%%%%%%%%%%%%%%%%%%%%
Black hole thermodynamics provides a unique window into quantum gravity, revealing profound connections between quantum information, statistical mechanics, and geometry \cite{PRD7-2333,
CMP43-199}. The development of extended phase space thermodynamics \cite{PRD52-4569,CQG17-399,
CPL23-1096,CQG26-195011,PRD84-024037,CQG34-063001,PRL132-021401,PRL132-191401} and holographic frameworks \cite{PRL127-091301,JHEP0822174,CQG39-075019,PRL130-181401,JHEP0623105,JHEP0823142, 2507.01010} has further enriched this field, leading to new insights into heat engines \cite{CQG31-205002}, complexity \cite{PRL126-101601,JHEP0521226}, entropy bounds \cite{PRL130-121501, PRL131-241401,PRL133-181501} and gravitational phase transitions \cite{CMP87-577,JHEP0712033,PRD88-121502,PRL118-021301,2505.24148,2506.10808,2507.10028}. Despite these advances, a fundamental challenge remains: to identify universal properties and classification schemes that transcend the details of specific black hole solutions.

Topology has recently emerged as a powerful tool to address this challenge, offering a way to categorize black hole phases based on global, quantitative invariants \cite{PRL129-191101,
PRD110-L081501,PRD111-L061501} \footnote{For a recent systematic exposition of the thermodynamic topology approach, including the classification of black hole phase states, Davies-type transition points, and critical points, please see Ref. \cite{2508.01614}.}. By treating solutions as topological defects in the thermodynamic parameter space, this approach has successfully classified black holes into distinct topological classes. Initial work identified three categories based on topological numbers \cite{PRL129-191101}, later refined into four broader classes via asymptotic thermodynamic behavior analysis \cite{PRD110-L081501}. Most recently, this framework was expanded to include five topological classes and two subclasses \cite{PRD111-L061501} \footnote{A broader set of illustrative examples can be found in Refs. \cite{PRD107-024024,PRD107-064023,PRD107-084002,EPJC83-365,EPJC83-589,PRD108-084041,JHEP0624213,
PLB856-138919,PDU46-101617,EPJC84-1294,CQG42-125007,PLB860-139163,PLB865-139482,EPJC85-828,2510.20164}, which collect some of the most recent representative developments in this area. The applicability of this topological framework also extends to the analysis of light rings \cite{PRL119-251102,
PRL124-181101,PRD102-064039,PRD103-104031,PLB858-139052,PLB868-139742} and timelike circular orbits \cite{PRD107-064006,JCAP0723049,PRD108-044077,CQG42-025020}, where it has consistently demonstrated its utility.}.

Despite these successful classifications, a natural and critical question remains: Is this topological classification complete? In this work, we demonstrate that it is not. Through a detailed investigation of higher-dimensional, multiply rotating Kerr-AdS black holes in odd dimensions, we have discovered a previously unrecognized topological subclass: the $\widetilde{W}^{1+}$ subclass. This new subclass exhibits thermodynamic stability properties that defy the existing classification, necessitating an expansion of the current topological framework.

The remaining part of this paper is organized as follows. In Sec. \ref{II}, we give a brief review of the thermodynamic topological approach \cite{PRD110-L081501}. To facilitate a direct comparison with the new topological subclass presented in Sec. \ref{IV}, we begin in Sec. \ref{III} by providing a systematic overview of the established properties characterizing the five known topological classes and two subclasses. In Sec. \ref{IV}, we present the novel $\widetilde{W}^{1+}$ subclass and elucidate its distinct characteristics through a comparative analysis. Finally, in Sec. \ref{V}, we present our conclusions and discuss the implications of our work,  where we also propose a candidate for a potential new topological subclass, $\widetilde{W}^{1-}$, to be explored in future studies.

%%%%%%%%%%%%%%%%%%%%%%%%%%%%%%%%%%%%%%%%%%%%%%%%%%%%%%%%%%%%%%%%%%%%%%%%%%%%
\section{A Brief review of the thermodynamic topological approach}\label{II}
%%%%%%%%%%%%%%%%%%%%%%%%%%%%%%%%%%%%%%%%%%%%%%%%%%%%%%%%%%%%%%%%%%%%%%%%%%%%
In this section, we provide a brief review of the thermodynamic topological approach. We begin by defining the generalized off-shell Helmholtz free energy \cite{PRD110-L081501}:
\be\label{FE}
\mathcal{F} = M -\frac{S}{\tau} \, .
\ee
Here, $M$ and $S$ are the black hole mass and Bekenstein-Hawking entropy, respectively. The parameter $\tau$ represents the inverse temperature of a cavity enclosing the black hole, placing the system in an off-shell state. It is crucial to note that the on-shell condition is uniquely achieved when $\tau$ equals the inverse Hawking temperature, i.e., $\tau = \beta = 1/T$. At this point, the generalized off-shell free energy $\mathcal{F}$ coincides with the standard Helmholtz free energy $F = M -TS$ \cite{PRD15-2752,PRD33-2092,PRD105-084030,PRD106-106015}.

Following Ref. \cite{PRD111-L061501}, the two-component vector field
\be
\phi = \big(\phi^{r_h}, \phi^\Theta \big) = \left(\frac{\p\mathcal{\hat{F}}}{\p r_h}, \frac{\p \mathcal{\hat{F}}}{\p\Theta}\right).
\ee
can be defined using the modified free energy $\mathcal{\hat{F}} = \mathcal{F} + 1/\sin\Theta$.  Here, $r_h$ denotes the event horizon radius, and $\Theta$ is an auxiliary parameter with a domain of $[0,+\infty]$. The condition $\phi^{r_h} = 0$ identifies black hole states as zero points (or defects) of the vector field. In addition, a key feature is the behavior of the component $\phi^\Theta$, which diverges at the boundaries $\Theta = 0$ and $\Theta = \pi$. This divergence signifies that the vector field points outward at both of these points.

One can employ Duan's theory of $\phi$-mapping topological currents \cite{SS9-1072,NPB514-705,
PRD61-045004} to define a conserved topological current. The construction begins with the coordinates $x^\nu = (\tau, r_h, \Theta)$ and the corresponding derivatives $\partial_\nu$. The normalized vector field $n^a$ is defined by its components:
$n^r = \phi^{r_h}/||\phi||$ and $n^\Theta = \phi^{\Theta}/||\phi||$. The topological current is then given by:
\be\label{jmu}
j^{\mu}=\frac{1}{2\pi}\epsilon^{\mu\nu\rho}\epsilon_{ab}\partial_{\nu}n^{a}\partial_{\rho}n^{b}, \qquad \mu,\nu,\rho=0,1,2.
\ee
By construction, this current is identically conserved, satisfying $\partial_{\mu}j^{\mu} = 0$. The topological current can be reformulated using the Jacobian determinant $J^{\mu}(\phi/x)$ as:
\be
j^{\mu}=\delta^{2}(\phi)J^{\mu}\Big(\frac{\phi}{x}\Big)\, .
\ee
This $\delta$-function representation makes it evident that $j^{\mu}$ is non-vanishing only at the isolated points where $\phi^a(x_i) = 0$. The total topological number $W$ in a parameter region $\Sigma$ is the integral of the temporal component:
\be
W = \int_{\Sigma}j^{0}d^2x \, .
\ee
This global invariant reduces to a sum of local winding numbers $w_i$ at each zero point:
\be
W = \sum_{i=1}^{N}\beta_{i}\eta_{i} = \sum_{i=1}^{N}w_{i} \, .
\ee
Here, the Hopf index $\beta_i$ counts the number of loops the field $\phi^a$ makes in its internal space as $x^{\mu}$ traverses a path around the zero point $z_i$, while the Brouwer degree $\eta_{i} = \mathrm{sign}(J^{0}({\phi}/{x})_{z_i})=\pm 1$ specifies the orientation of the field mapping at the zero. Their product defines the winding number $w_i$ for the $i$th zero point. Crucially, the winding number is an intrinsic property of the zero point itself; any two distinct closed curves $\Sigma_1$ and $\Sigma_2$ enclosing the same zero point must yield the same winding number. Furthermore, the topological number $W$ vanishes identically for any region $\Sigma$ that contains no zero points of $\phi$.

A key insight from this topological framework is the assignment of stability: locally stable black hole states possess a local topological number $w = +1$, in contrast to unstable states, which have $w = -1$. Consequently, the global topological number $W$ serves as a robust classifier for different types of black holes. This provides a powerful global taxonomy based solely on topological invariants.

%%%%%%%%%%%%%%%%%%%%%%%%%%%%%%%%%%%%%%%%%%%%%%%%%%%%%%%%%%%%%%%%%%%%%%%%%%%%
\section{Five topological classes and two topological subclasses}\label{III}
%%%%%%%%%%%%%%%%%%%%%%%%%%%%%%%%%%%%%%%%%%%%%%%%%%%%%%%%%%%%%%%%%%%%%%%%%%%%
This section is dedicated to a systematic overview of the currently known five topological classes and two subclasses, detailing their established properties. This foundation is essential for the direct comparison with the new topological subclass that will be introduced in the following section.

According to our previous work \cite{PRD111-L061501}, the known topological classification comprises the following classes and subclasses:
\be\label{7TC}
W^{1-} \, , ~~ W^{0+} \, , ~~ W^{0-} \, , ~~ W^{1+} \, , ~~ W^{0-\leftrightarrow 1+} \, , ~~ \overline{W}^{1+} \, , ~~ \hat{W}^{1+} \, ,
\ee
which correspond to distinct asymptotic behaviors of the inverse temperature $\beta(r_h)$:
\bea
&W^{1-}&:~  \beta(r_{m})=0\, ,\quad\;\; \beta(\infty)=\infty \, , \label{b1-} \\
&W^{0+}&:~  \beta(r_{m})=\infty \, ,\quad \beta(\infty)=\infty \, , \label{b0+}   \\
&W^{0-}&:~  \beta(r_{m})=0\, ,\quad\;\; \beta(\infty)=0 \, , \label{b0-} \\
&W^{1+},~\hat{W}^{1+}&:~  \beta(r_{m})=\infty \, ,\quad \beta(\infty)=0 \, , \label{b1+} \\
&W^{0-\leftrightarrow 1+},~\overline{W}^{1+}&:~ \beta(r_m) = {\rm fixed~temperature}\, , ~ \beta(\infty) = 0 \, , \qquad \label{b01}
\eea
where $r_m$ denotes the minimal possible event horizon radius of the black hole, a value that can either be zero or nonzero. Consider the case of a Reissner-Nordstr\"om (RN) black hole with fixed charge $Q$: here, $r_m$ equals the mass at extremality, yielding $r_m = M = Q = r_e$. By contrast, a Schwarzschild black hole has $r_m = 0$.

We now examine the asymptotic behavior of $\phi$ at the boundary corresponding to Eqs. (\ref{b1-})-(\ref{b01}), which is described by the contour $C=I_1\cup I_2\cup I_3\cup I_4$, where $I_{1}=\{r_{h}=\infty,\;\Theta\in(0, \pi)\}$, $I_{2}=\{r_{h}=(\infty,\;r_{m}),\;\Theta=\pi\}$, $I_{3}=\{r_{h}=r_{m},\;\Theta\in(\pi, 0)\}$, and $I_{4}=\{r_{h}=(r_{m},\;\infty),\;\Theta=0\}$. This contour encloses all possible parameter regions. Given that $\phi$ is defined to be orthogonal to $I_2$ and $I_4$ \cite{PRL129-191101,PRL124-181101}, we now analyze its asymptotic behavior along $I_1$ and $I_3$. We begin with the $r_h$ component. Using the first law, we express it as:
\be
\phi^{r_{h}}=\frac{\partial\hat{\mathcal{F}}}{\partial r_{h}}
= \frac{\partial S}{\partial r_{h}}\left(\frac{1}{\beta}-\frac{1}{\tau}\right)\, .
\ee
The cavity temperature $\tau$ is a fixed positive constant. Hence, for $\frac{\partial S}{\partial r_{h}} > 0$, the behavior of $\phi^{r_{h}}$ is dictated solely by $\beta$: it becomes positive as $\beta \to 0$ and negative as $\beta \to \infty$. Thus, near the boundaries $r_h \to r_m$ and $r_h \to \infty$, the vector $\phi$ points rightward or leftward, with some inclination determined by $\phi^{\Theta}$.

\begin{table}[b]
\caption{Topological (sub)classes: Direction of $\phi^{r_h}$ arrows and corresponding topological numbers for the four segments. \label{TabI}}
\begin{tabular}{c|c|c|c|c|c}\hline\hline
Topological (sub)classes     & $I_{1}$ &$I_{2}$ & $I_{3}$ & $I_{4}$ &$W$ \\ \hline
$W^{1-}$      & $\leftarrow$ & $\uparrow$& $\rightarrow$ & $\downarrow$ &-1 \\
$W^{0+}$   & $\leftarrow$   &$\uparrow$ & $\leftarrow$     &$\downarrow$ &0 \\
$W^{0-}$   & $\rightarrow$ &$\uparrow$ & $\rightarrow$ & $\downarrow$ &0 \\
$W^{1+}, ~\overline{W}^{1+}, ~\hat{W}^{1+}$      &  $\rightarrow$   & $\uparrow$& $\leftarrow$ & $\downarrow$ &+1\\
$W^{0-\leftrightarrow 1+}$ (when $W = 0$)  & $\rightarrow$ &$\uparrow$ & $\rightarrow$ & $\downarrow$ &0 \\
$W^{0-\leftrightarrow 1+}$ (when $W = 1$) &  $\rightarrow$   & $\uparrow$& $\leftarrow$ & $\downarrow$ &+1 \\
\hline\hline
\end{tabular}
\end{table}

Table \ref{TabI} summarizes the directions of the $\phi^{r_h}$ arrows for all four segments in each of the seven aforementioned topological (sub)classes defined in Eq. (\ref{7TC}), along with their corresponding topological numbers.

%%%%%%%%%%%%%%%%%%%%%%%%%%%%%%%%%%%%%%%%%%%%%%%%%%%%%%%%%%%%%%%%%%%%%%%%%%%%%%
\begin{figure*}[t]
\subfigure[]
{\label{W1-}
\includegraphics[width=0.22\textwidth]{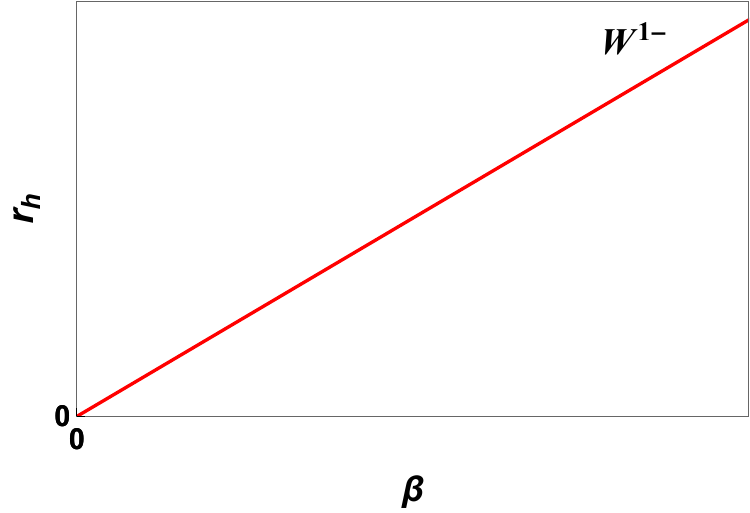}}
\subfigure[]
{\label{W0+}
\includegraphics[width=0.22\textwidth]{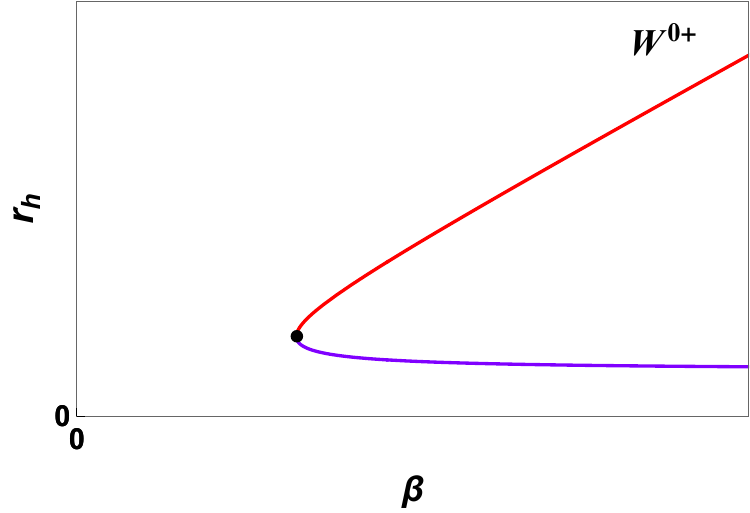}}
\subfigure[]
{\label{W0-}
\includegraphics[width=0.22\textwidth]{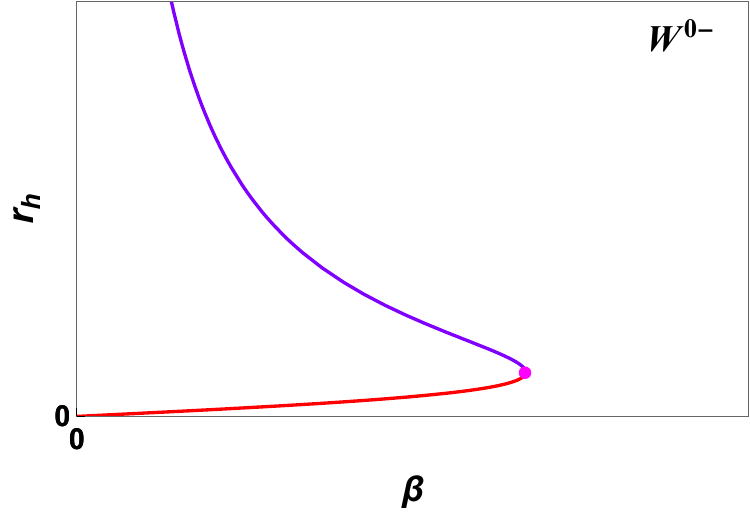}}
\subfigure[]
{\label{W1+}
\includegraphics[width=0.22\textwidth]{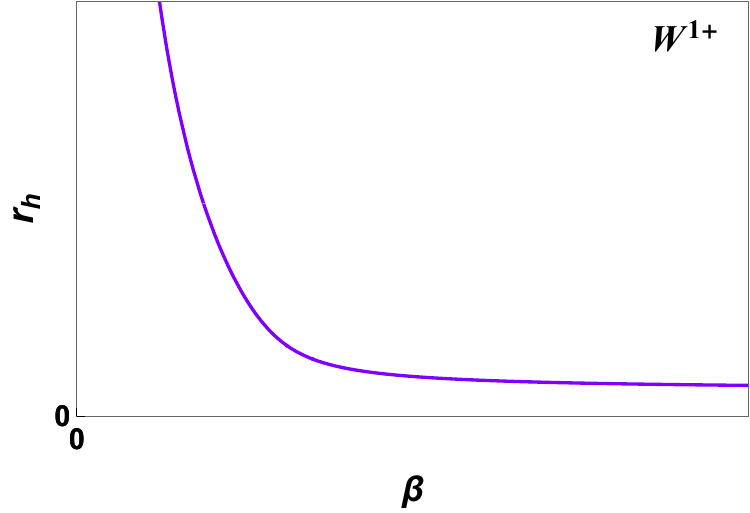}}
\subfigure[]
{\label{W0-1+}
\includegraphics[width=0.22\textwidth]{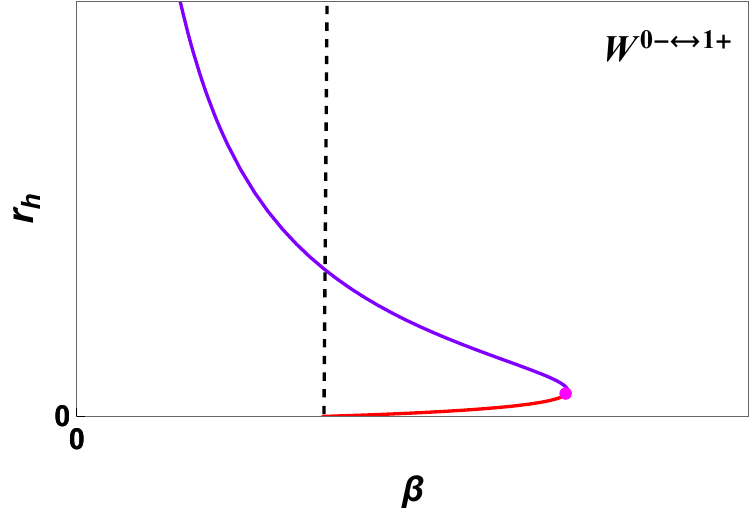}}
\subfigure[]
{\label{bW1+}
\includegraphics[width=0.22\textwidth]{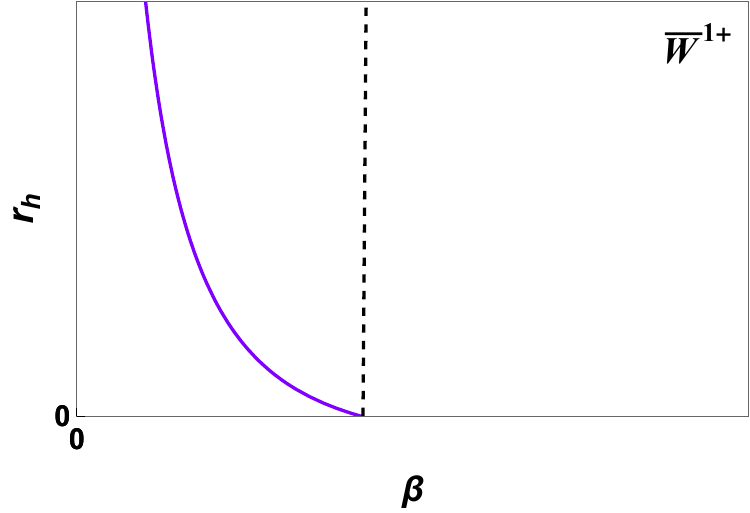}}
\subfigure[]
{\label{hW1+}
\includegraphics[width=0.22\textwidth]{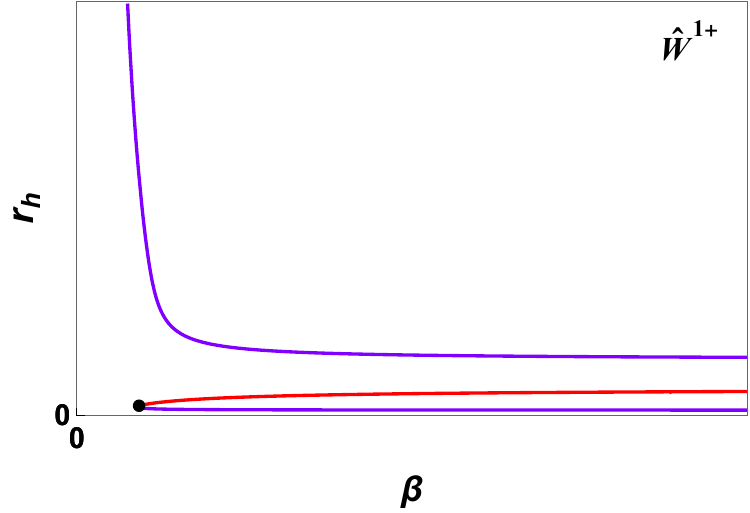}}
\caption{The zero points of $\phi^{r_h}$ are shown in the $r_h-\beta$ plane for $W^{1-}$, $W^{0+}$, $W^{0-}$, $W^{1+}$, $W^{0-\leftrightarrow 1+}$, $\overline{W}^{1+}$, $\hat{W}^{1+}$ (sub)classes, respectively. The red line corresponds to a thermodynamically unstable black hole branch with $w = -1$, while the purple line corresponds to a thermodynamically stable black hole branch with $w = 1$. The black dot represents the generation point (GP) within the degenerate point (DP), and the pink dot represents the annihilation point (AP) within the DP. (a) Typical case: four-dimensional Schwarzschild black hole. (b) Typical case: four-dimensional Kerr black hole. (c) Typical case: four-dimensional Schwarzschild-AdS black hole. (d) Typical case: four-dimensional Kerr-AdS black hole. (e) Typical case: four-dimensional static two-charge AdS black hole when $q_1 < q_{1c} = 3/(8\pi Pq_2)$. (f) Typical case: four-dimensional static two-charge AdS black hole when $q_1 \ge q_{1c} = 3/(8\pi Pq_2)$. (g) Typical case: four-dimensional dyonic AdS black hole.
\label{fig1}}
\end{figure*}
%%%%%%%%%%%%%%%%%%%%%%%%%%%%%%%%%%%%%%%%%%%%%%%%%%%%%%%%%%%%%%%%%%%%%%%%%%%%%%

\onecolumngrid
\begin{center}
\begin{table}[t]
\caption{Thermodynamic properties of the black hole states for the seven topological (sub)classes of $W^{1-}$, $W^{0+}$, $W^{0-}$, $W^{1+}$, $W^{0-\leftrightarrow 1+}$, $\overline{W}^{1+}$, and $\hat{W}^{1+}$, respectively. \label{TabII}}
\begin{tabular}{c|c|c|c|c|c|c}\hline\hline
Topological (sub)classes  & Innermost & Outermost & Low $T$ ($\beta\to\infty$) & High $T$ ($\beta\to 0$) & DP & $W$ \\ \hline
$W^{1-}$     & unstable & unstable & unstable large & unstable small  & in pairs & $-1$ \\
$W^{0+}$    & stable     & unstable & unstable large + stable small    & no & one more GP & $0$ \\
$W^{0-}$     & unstable & stable     & no & unstable small + stable large  & one more AP & $0$ \\
$W^{1+}$    & stable     & stable     & stable small  & stable large & in pairs & $+1$\\
$W^{0-\leftrightarrow 1+}$     & unstable & stable & no & stable large  & one more AP & $0$ or $+1$ \\
$\overline{W}^{1+}$    & stable  & stable   & no & stable large & in pairs & $+1$ \\
$\hat{W}^{1+}$    & stable     & stable     & unstable small+two stable small  & stable large & one more GP & $+1$\\
\hline\hline
\end{tabular}
\end{table}
\end{center}
\twocolumngrid

Next, we discuss the typical black hole solutions that fall into each of these seven topological (sub)classes.

In the $W^{1-}$ class, the four-dimensional Schwarzschild black hole can be considered the most prototypical member \cite{PRL129-191101,PRD110-L081501}. For a fixed value of $\tau$, only one black hole state characterized by a negative heat capacity exists. The local winding number corresponding to this state is $-1$, thereby agreeing with the global topological number of $W = -1$.

Within the $W^{0+}$ topological class, the four-dimensional Kerr black hole is a canonical example \cite{PRD107-024024,PLB860-139163}. The introduction of a rotation parameter markedly alters the thermodynamics: small black hole state with a nonzero rotation parameter exhibits positive heat capacity, while its large counterpart maintains a negative heat capacity. This system exhibits a generation point at a critical value $\beta_*$ \cite{PRD107-024024}, below which no black hole states exist, yielding a trivial topology. At higher $\beta$, two distinct phases coexist: a stable small black hole state and an unstable large black hole state. Given their opposite stability characters, these phases carry winding numbers of opposite signs, which cancel out, resulting in a vanishing topological number.

Table \ref{TabI} shows that the topological number is zero for both the $W^{0-}$ and $W^{0+}$ classes; nevertheless, the black holes in these categories demonstrate markedly distinct asymptotic thermodynamic behaviors at small and large values of the horizon radius $r_h$. The four-dimensional Schwarzschild-AdS black hole serves as a representative example within the $W^{0-}$ class \cite{PRD107-084002,PRD110-L081501}. In contrast to its Schwarzschild counterpart, the presence of a nonzero negative cosmological constant here introduces significant modifications: small AdS black holes exhibit negative heat capacity, whereas large AdS black holes possess positive heat capacity. This system features an annihilation point at a critical value $\beta_*$. For $\beta > \beta_*$, the topology is trivial due to the absence of black hole states. Conversely, at small $\beta$, two coexisting states emerge: a thermodynamically unstable small black hole and a stable large one.

The $W^{1+}$ class is characterized by a topological number of $1$, which reveals an additional locally stable black hole state in the phase space for a fixed $\tau$. The four-dimensional Kerr-AdS black hole is a concrete realization of this case. The key effect of the negative cosmological constant is to introduce a new stable phase, supplementing the two-phase structure (comprising one unstable large black hole state and one stable small black hole state) of the asymptotically flat Kerr black hole. This explains the resulting topological number of $W = 1$. In the regime of a small cosmological constant $\Lambda$, three states coexist (two stable and one unstable) at intermediate $\tau$, whereas a large $\Lambda$ eliminates the unstable branch, resulting in a single stable black hole for all $\tau$. Notably, the topological number remains robustly fixed at $1$ and is insensitive to variations in $\tau$ or $\Lambda$.

In contrast to the four topological classes above, each characterized by a single topological invariant, the $W^{0-\leftrightarrow 1+}$ class is the first known to simultaneously possess two distinct topological numbers: $0$ and $1$. This phenomenon arises from a novel thermodynamic topological phase transition that occurs in black holes of this class as temperature varies. When the electric charge parameter satisfies $q_1 < q_{1c} = 3/(8\pi Pq_2)$, the four-dimensional static two-charge AdS black hole constitutes a typical example of the $W^{0-\leftrightarrow 1+}$ class \cite{JHEP0624213}. Asymptotically, the thermodynamic behavior is characterized by a phase transition: the black hole state is absent at low temperatures ($\beta \to \infty$), giving way to a stable large black hole phase in the high-temperature limit ($\beta \to 0$). Remarkably, the asymptotic thermodynamic behaviors of the four-dimensional static two-charge AdS black hole are fundamentally distinct from those of the corresponding RN-AdS black hole. This divergence compellingly argues for the necessity and demonstrates the significant potential of exploring the phase structure of charged AdS black holes in the framework of gauged supergravity. However, currently there is no known related work, and therefore much has to be done in this direction.

Then, we discuss the black hole solutions within the $\overline{W}^{1+}$ and $\hat{W}^{1+}$ subclasses and their asymptotic thermodynamic behavior. Black hole solutions in the $\overline{W}^{1+}$ and $\hat{W}^{1+}$ subclasses have a topological number of $1$, identical to that of the $W1+$ class. However, their asymptotic thermodynamic behaviors are unique.  The four-dimensional static two-charge AdS black hole, under the condition $q_1 \ge q_{1c} = 3/(8\pi Pq_2)$, is a canonical member of the $\overline{W}^{1+}$ subclass \cite{JHEP0624213}. On the other hand, the four-dimensional dyonic AdS black hole serves as a representative for the $\hat{W}^{1+}$ subclass \cite{EPJC84-1294}. In the high-temperature limit ($\beta \to 0$), the black holes in the $\overline{W}^{1+}$, $\hat{W}^{1+}$, and $W^{1+}$ classes all exhibit identical thermodynamic behavior: the presence of a single, thermodynamically stable large black hole state. However, in the low-temperature limit ($\beta \to \infty$), their behaviors diverge significantly. For the $\overline{W}^{1+}$ subclass, no black hole state exists. In contrast, the $\hat{W}^{1+}$ subclass exhibits three possible states: one thermodynamically unstable small black hole and two thermodynamically stable small black holes.

Last but not least, Fig. \ref{fig1} and Table \ref{TabII} serve as a comprehensive summary of the characteristics of the seven known topological (sub)classes, thus providing a foundation for understanding their thermodynamic properties.

%%%%%%%%%%%%%%%%%%%%%%%%%%%%%%%%%%%%%%%%%%%%%%%%%%%%%%%%%
\section{Newly identified topological subclass: $\widetilde{W}^{1+}$}\label{IV}
%%%%%%%%%%%%%%%%%%%%%%%%%%%%%%%%%%%%%%%%%%%%%%%%%%%%%%%%%
This section shifts focus to the novel topological subclass identified in higher-odd-dimensional, multiply-rotating Kerr-AdS black holes, offering a comprehensive examination of their asymptotic thermodynamic behavior.

In the generalized Boyer-Lindquist coordinates, the metric of the $d$-dimensional, multiply-rotating Kerr-AdS black holes is given by \cite{PRL93-171102,
JHEP0615096}:
\be
ds^2 = d\gamma^2 +\frac{2m}{U}\omega^2 +\frac{Udr^2}{F -2m} +d\Omega^2 \, ,
\ee
where
\bea
d\gamma^2 &=& -\frac{W\rho^2}{l^2}dt^2 +\sum_{i=1}^N\frac{r^2 +a_i^2}{\Xi_i}\mu_i^2d\varphi_i^2 \, , \nn \\
d\Omega^2 &=& \sum_{i=1}^{N+\epsilon}\frac{r^2 +a_i^2}{\Xi_i}d\mu_i^2 -\frac{1}{W\rho^2}\left(\sum_{i=1}^{N+\epsilon}\frac{r^2 +a_i^2}{\Xi_i}\mu_id\mu_i \right)^2 \, ,
\nn \\
\omega &=& Wdt -\sum_{i=1}^N\frac{a_i\mu_i^2d\varphi_i}{\Xi_i} \, , \nn
\eea
in which $\rho^2 = r^2 +l^2$, and
\bea
&&W =  \sum_{i=1}^{N+\epsilon}\frac{\mu_i^2}{\Xi_i} \, , \quad U = r^\epsilon\sum_{i=1}^{N+\epsilon}\frac{\mu_i^2}{r^2 +a_i^2}\prod_j^N(r^2 +a_j^2) \, ,\nn \\
&&\Xi_i = 1 -\frac{a_i^2}{l^2} \, , ~\quad F = \frac{r^{\epsilon-2}\rho^2}{l^2}\prod_{i=1}^N(r^2 +a_i^2) \, . \nn
\eea
Here, $m$ is the mass parameter, $a_i$ are the independent rotation parameters, and $l$ is the AdS radius. To uniformly treat spacetimes of both even and odd dimensionality, we introduce the parameter $\epsilon$, where $\epsilon = 1$ for even dimensions and $\epsilon = 0$ for odd ones. The dimension $d$ is consequently expressed as
\be
d = 2N +1 +\epsilon \, .
\ee
Furthermore, we impose the convenient convention that $a_{N+1} = 0$ in even dimensions. Lastly, the coordinates $\mu_i$ are related by the constraint
\be
\sum_{i=1}^{N+\epsilon}\mu_i^2 = 1 \, .
\ee
The spacetime typically features $N$ independent angular momenta $J_i$, parameterized by $N$ rotation parameters $a_i$. Specifically, the expressions for the mass $M$, the angular momenta $J_i$, and the horizon angular velocities $\Omega_i$ are as follows \cite{CQG22-1503}:
\be\ba
&M = \frac{m\mathcal{A}_{d-2}}{4\pi\left(\prod_j\Xi_j \right)}\left(\sum_{i=1}^N\frac{1}{\Xi_i} -\frac{1 -\epsilon}{2} \right) \, , \\
&J_i = \frac{ma_i\mathcal{A}_{d-2}}{4\pi\Xi_i\left(\prod_j\Xi_j \right)} \, , \qquad \Omega_i = \frac{a_i}{r_h^2 +a_i^2}\left(1 +\frac{r_h^2}{l^2} \right) \, ,
\ea\ee
while the temperature $T$, horizon area $A$, and entropy $S$ are given by
\be\ba
&T = \frac{1}{2\pi}\left[\left(r_h +\frac{r_h^3}{l^2} \right)\sum_{i=1}^N\frac{1}{r_h^2 +a^2} -\frac{1}{r_h}\left(\frac{1}{2} -\frac{r_h^2}{2l^2} \right)^\epsilon\right] \, , \\
&A = \frac{\mathcal{A}_{d-2}}{r_h^{1-\epsilon}}\prod_{i=1}^N\frac{r_h^2 +a_i^2}{\Xi_i} \, , \qquad S = \frac{A}{4} \, .
\ea\ee
The horizon radius $r_h$ is given by the largest root of $F -2m = 0$, while the area of the unit $(d-2)$-sphere is $\mathcal{A}_{d-2} = 2\pi^{[(d-1)/2]}/\Gamma[(d-1)/2]$.

Given that the asymptotic thermodynamic behavior observed in five-, seven-, and nine-dimensional multi-rotating Kerr-AdS black holes is homologous, they all fall into the same thermodynamic topological class. To prevent unnecessary repetition in the main body of the text, a detailed discussion is reserved for the five-dimensional case as a representative example. The corresponding analyses for the seven- and nine-dimensional configurations are presented in Appendices \ref{AppA} and \ref{AppB}, respectively.

It is easily observed that the Hawking temperature of the five-dimensional doubly-rotating Kerr-AdS black hole diverges not only in the limit $r_h\to r_m$, but also as $r_h\to\infty$. It follows that the inverse temperature $\beta$ exhibits the behavior
\be
\beta(r_m) = 0 \, , \qquad \beta(\infty) = 0
\ee
at these asymptotic boundaries.

%%%%%%%%%%%%%%%%%%%%%%%%%%%%%%%%%%%%%%%%%%%%%%%%%%%%%%%%%%%%%%%%%%%%%%%%%%%%%%%%%%%%%%%%%%
\begin{figure}[t]
\centering
\includegraphics[width=0.35\textwidth]{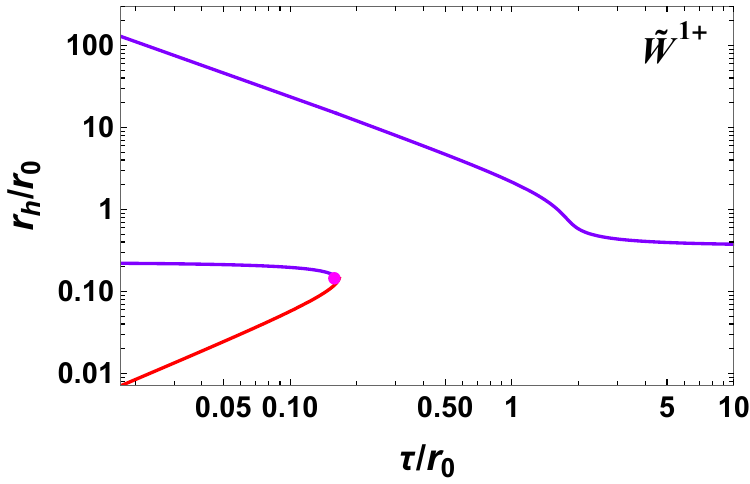}
\caption{Zero points of the vector $\phi^{r_h}$ on the $r_h$-$\tau$ plane for the five-dimensional doubly-rotating Kerr-AdS black hole, with parameters $a_1/r_0 = 0.5$, $a_2/r_0 = 1/3$, and $l/r_0 = 1$. Here, the thermodynamically unstable ($w = -1$, red line) and stable ($w = +1$, purple line) branches intersect at the AP (pink dot). Counting the number of stable and unstable states reveals two of the former and one of the latter, which yields a topological number $W = -1 +1 +1 = 1$.
\label{fig2}}
\end{figure}
%%%%%%%%%%%%%%%%%%%%%%%%%%%%%%%%%%%%%%%%%%%%%%%%%%%%%%%%%%%%%%%%%%%%%%%%%%%%%%%%%%%%%%%%%%

Substituting the relation $\mathcal{A}_3 = 2\pi^2$ into the definition of the generalized off-shell Helmholtz free energy (\ref{FE}) simplifies the expression for $\mathcal{F}$, resulting in
\bea
\mathcal{F} &=& \frac{\pi(r_h^2 +l^2)(2\Xi_1 +2\Xi_2 -\Xi_1\Xi_2)}{8r_h^2l^2}\prod_{i=1}^2\frac{r_h^2 +a_i^2}{\Xi_i^2} \nn \\
&&-\frac{\pi^2}{2r_h\tau}\prod_{i=1}^2\frac{r_h^2 +a_i^2}{\Xi_i} \, .
\eea
Therefore, the components of the vector $\phi$ are
\bea
\phi^{r_h} &=& \frac{\pi r_h^4(2r_h^2 +a_1^2 +a_2^2)(2\Xi_1 +2\Xi_2 -\Xi_1\Xi_2)}{4r_h^3l^2\Xi_1^2\Xi_2^2} \nn \\
&&+\frac{(r_h^2 -a_1a_2)(r_h^2 +a_1a_2)(2\Xi_1 +2\Xi_2 -\Xi_1\Xi_2)}{4r_h^3\Xi_1^2\Xi_2^2} \qquad \nn \\
&&-\frac{\pi^2(3r_h^4 +r_h^2a_1^2 +r_h^2a_2^2 -a_1^2a_2^2)}{2r_h^2\tau\Xi_1\Xi_2} \, , \\
\phi^{\Theta} &=& -\cot\Theta\csc\Theta \, .
\eea
Thus, one can compute the zero point of the vector field $\phi^{r_h}$ as
\be
\tau = \frac{2\pi r_hl^2\Xi_1\Xi_2(3r_h^4 +r_h^2a_1^2 +r_h^2a_2^2 -a_1^2a_2^2)}{[2r_h^6 +r_h^4(a_1^2 +a_2^2 +l^2) -a_1^2a_2^2l^2](2\Xi_1 +2\Xi_2 -\Xi_1\Xi_2)} \, . \qquad
\ee

Figure \ref{fig2} shows the zero points of the vector $\phi^{r_h}$ on the $r_h$-$\tau$ plane for a five-dimensional doubly rotating Kerr-AdS black hole with a generic configuration of non-equal rotation parameters ($a_1/r_0 = 0.5$, $a_2/r_0 = 1/3$, $l/r_0 = 1$). The equal-rotation case ($a_1 = a_2$) is omitted, as the qualitative results remain unchanged. Here, $r_0$ is an arbitrary length scale set by the cavity enclosing the black hole. The thermodynamic phase structure consists of three branches: a stable large black hole branch ($w = +1$), a stable small black hole branch ($w = +1$), and an unstable small black hole branch ($w = -1$). The two small black hole branches merge and annihilate at the critical point, marked in pink. This confirms the coexistence of three distinct black hole states within a certain temperature range. The total topological number, given by the sum $W = -1 +1 +1 = 1$, is consistent with this three-state picture. Furthermore, by comparing Fig. \ref{fig2} with Fig. \ref{fig1}, we can readily observe that the asymptotic thermodynamic behavior of the five-dimensional doubly rotating Kerr-AdS black hole differs from all seven known thermodynamic topology (sub)classes. Despite this, its topological number is identical to that of the $W^{1+}$ class, both being $1$. We therefore propose that it belongs to a previously unidentified topological subclass.

Our analysis of the asymptotic thermodynamic behavior of this novel topological subclass will proceed as follows. First, we examine the systematic ordering within this new subclass. This system features a minimum of three black hole states: two with positive heat capacity and a winding number of $1$, and one with negative heat capacity and a winding number of $-1$. Any  additional states must emerge in pairs, consisting of states with oppositely signed winding numbers. The winding number sequence for these black holes does not follow a simple alternation. Instead, it begins with two specific states corresponding to the sequence $[-, +]$. All subsequent states, starting from the third one, follow a distinct pattern: they form a block that starts with a positive winding number $(+)$, ends with a positive winding number $(+)$, and contains pairs of $[-, +]$ in between, resulting in a subsequence like $[+, -, +, -, ..., +]$. This structure ensures that the innermost black hole has $w = -1$ (unstable) and the outermost has $w = +1$ (stable), leading to the topological classification $[-, +]$ based on these endpoints.

Then, we turn our attention to the asymptotic thermodynamic behavior of this new topological subclass. In the low-temperature limit ($\beta\to\infty$), the system admits only a single, thermodynamically stable small black hole. In contrast, the high-temperature limit ($\beta\to 0$) features a rich phase structure comprising three distinct states: one stable large black hole, one stable small black hole, and one unstable small black hole. Owing to its distinct thermodynamic properties and following the consistent convention of progressing from small to large black hole states, we propose the nomenclature $\widetilde{W}^{1+}$ for this new topological subclass.

\onecolumngrid

%%%%%%%%%%%%%%%%%%%%%%%%%%%%%%%%%%%%%%%%%%%%%%%%%%%%%%%%%%%%%%%%%%%%%%%%%%%%%%%%%%%%%%%%
\begin{center}
\begin{table}[h]
\caption{Thermodynamic properties of black hole states of the new topological subclass $\widetilde{W}^{1+}$.  \label{TabIII}}
\begin{tabular}{c|c|c|c|c|c|c}\hline
Topological subclass  & Innermost & Outermost & Low $T$ ($\beta\to\infty$) & High $T$ ($\beta\to 0$) & DP & $W$ \\ \hline
$\widetilde{W}^{1+}$     & unstable & stable & stable small & unstable small + stable small + stable large  & one more AP & $+1$ \\
\hline
\end{tabular}
\end{table}
\end{center}
%%%%%%%%%%%%%%%%%%%%%%%%%%%%%%%%%%%%%%%%%%%%%%%%%%%%%%%%%%%%%%%%%%%%%%%%%%%%%%%%%%%%%%%%

%%%%%%%%%%%%%%%%%%%%%%%%%%%%%%%%%%%%%%%%%%%%%%%%%%%%%%%%%%%%%%%%%%%%%%%%%%%%%%%%%%%%%%%%
\begin{center}
\begin{table}[h]
\caption{Proposed thermodynamic properties of black hole states of the possible new topological subclass $\widetilde{W}^{1-}$.  \label{TabIV}}
\begin{tabular}{c|c|c|c|c|c|c}\hline
Topological subclass  & Innermost & Outermost & Low $T$ ($\beta\to\infty$) & High $T$ ($\beta\to 0$) & DP & $W$ \\ \hline
$\widetilde{W}^{1-}$     & stable & unstable & stable small + unstable small + unstable large & unstable small  & one more GP & $-1$ \\
\hline
\end{tabular}
\end{table}
\end{center}
%%%%%%%%%%%%%%%%%%%%%%%%%%%%%%%%%%%%%%%%%%%%%%%%%%%%%%%%%%%%%%%%%%%%%%%%%%%%%%%%%%%%%%%%

\twocolumngrid

%%%%%%%%%%%%%%%%%%%%%%%%%%%%%%%%%%%%%%%%%%%%%%
\section{Conclusions and discussions}\label{V}
%%%%%%%%%%%%%%%%%%%%%%%%%%%%%%%%%%%%%%%%%%%%%%

By analyzing higher-odd-dimensional, multiply rotating Kerr-AdS black holes as topological defects, we have discovered a previously overlooked thermodynamic phenomenon, leading to the identification of a novel topological subclass: the $\widetilde{W}^{1+}$ subclass. This finding compels us to move beyond the established classification of five classes and two subclasses. The essence of this new subclass is captured in Table \ref{TabIII}, which summarizes its distinctive stability characteristics. What makes the $\widetilde{W}^{1+}$ subclass truly singular is its behavior at temperature extremes. At low temperatures, it hosts a unique configuration with only one stable state, namely a small black hole, which challenges our intuition about typical phase transitions. At high temperatures, it presents a rich tapestry of three coexisting states, including one stable large black hole, one stable small black hole, and one unstable small black hole. This work not only confirms that the thermodynamic landscape of black holes is richer than previously thought but also demonstrates the power of the topological approach in uncovering this hidden complexity.

The symmetry observed in the functions $(r_h,\beta)$ suggests a fascinating possibility: the existence of a thermodynamic topological subclass that would serve as the symmetric counterpart to the novel $\widetilde{W}^{1+}$ subclass we have presented. This hypothetical class would exhibit an inverse stability profile, effectively completing the symmetric picture. The future task of determining whether actual black hole solutions embody this phenomenology presents a compelling direction for research. We thus propose the existence of, and designate, this candidate subclass as $\widetilde{W}^{1-}$.

\begin{figure}[h]
\centering
\includegraphics[width=0.28\textwidth]{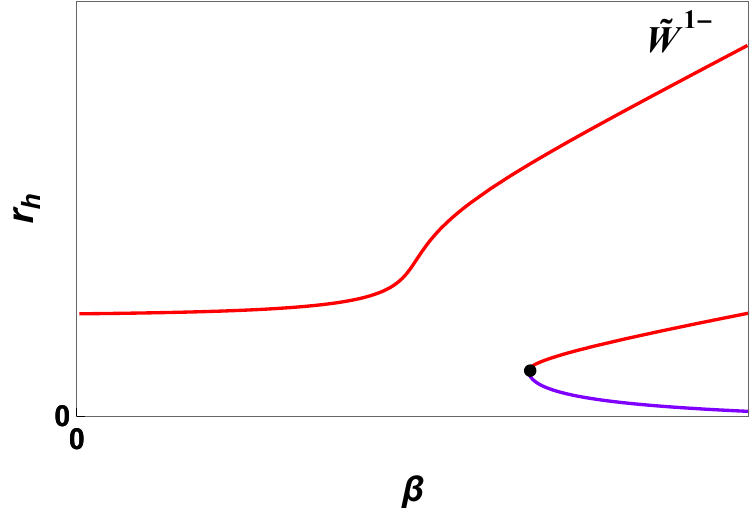}
\caption{The schematic diagram of the zero points of $\phi^{r_h}$ is shown in the $r_h-\beta$ plane for the possible new topological subclass $\widetilde{W}^{1-}$.
\label{fig3}}
\end{figure}

The proposed $\widetilde{W}^{1-}$ subclass, as illustrated in Fig. \ref{fig3} and summarized in Table \ref{TabIV}, would be characterized by an inverse stability profile compared to the $\widetilde{W}^{1+}$ subclass. Specifically, its low-temperature limit would be dominated by a single, stable large black hole state. In the high-temperature regime, the phase structure would consist of one stable small black hole, one unstable small black hole, and one stable large black hole. This configuration would complete the symmetric picture suggested by the functional analysis. The primary outstanding question for this candidate subclass is whether its thermodynamic structure can be realized by physical black hole solutions, a crucial focus for future investigations.

The confirmation of the $\widetilde{W}^{1+}$ subclass and the potential for this $\widetilde{W}^{1-}$ counterpart collectively demonstrate that the current topological classification scheme must be expanded to accommodate the newly discovered phenomenology. Our results provide a new foundation for classifying the thermodynamics of complex black objects and suggest that further unexplored topological classes may await discovery.

%%%%%%%%%%%%%%%%%%%%%%%%%%%%%%%%%%%%%%%%%%%%%%%%%%%
%%%%%%%%%%%%%%%%%%%%%%%%%%%%%%%%%%%%%%%%%%%%%%%%%%%%

\acknowledgments

This work is supported by the National Natural Science Foundation of China (NSFC) under Grants No. 12205243, No. 12375053.

\appendix

\onecolumngrid

%%%%%%%%%%%%%%%%%%%%%%%%%%%%%%%%%%%%%%%%%%%%%%%%%%%%%%%%%%%%%%%%%%%%%%%%%%%%%%%
\section{Seven-dimensional triple-rotating Kerr-AdS black hole case}\label{AppA}
%%%%%%%%%%%%%%%%%%%%%%%%%%%%%%%%%%%%%%%%%%%%%%%%%%%%%%%%%%%%%%%%%%%%%%%%%%%%%%%
In this appendix, we investigate the thermodynamic topological classification of the seven-dimensional, triple-rotating Kerr-AdS black hole. Upon making the substitutions $\mathcal{A}_5 = \pi^3$ in the definition of the generalized off-shell Helmholtz free energy (\ref{FE}), the expression for $\mathcal{F}$ reduces to the form
\be
\mathcal{F} = \frac{\pi^2(r_h^2 +l^2)[2(\Xi_1\Xi_2 +\Xi_1\Xi_3 +\Xi_2\Xi_3) -\Xi_1\Xi_2\Xi_3]}{16r_h^2l^2}\prod_{i=1}^3\frac{r_h^2 +a_i^2}{\Xi_i^2} -\frac{\pi^3}{4r_h\tau}\prod_{i=1}^3\frac{r_h^2 +a_i^2}{\Xi_i} \, .
\ee
The components of the vector $\phi$ can be easily calculated as:
\bea
\phi^{r_h} &=& \frac{\pi^2\Big[3r_h^8 +2r_h^6\Big(l^2 +\sum_{i=1}^3a_i^2 \Big) +r_h^4\Big(l^2\sum_{i=1}^3a_i^2 +a_1^2a_2^2 +a_1^2a_3^2 +a_2^2a_3^2 \Big) -l^2a_1^2a_2^2a_3^2\Big][2(\Xi_1\Xi_2 +\Xi_1\Xi_3 +\Xi_2\Xi_3) -\Xi_1\Xi_2\Xi_3]}{8r_h^3l^2\Xi_1^2\Xi_2^2\Xi_3^2} \nn \\
&&-\frac{\pi^3\Big[5r_h^6 +3r_h^4\sum_{i=1}^3a_i^2 +r_h^2(a_1^2a_2^2 +a_1^2a_3^2 +a_2^2a_3^2) -a_1^2a_2^2a_3^2\Big]}{4r_h^2\tau\Xi_1\Xi_2\Xi_3} \, ,  \\
\phi^{\Theta} &=& -\cot\Theta\csc\Theta \, .
\eea
By solving the equation $\phi^{r_h} = 0$, one can obtain
\be
\tau = \frac{2\pi r_hl^2\Xi_1\Xi_2\Xi_3\Big[5r_h^6 +3r_h^4\sum_{i=1}^3a_i^2 +r_h^2(a_1^2a_2^2 +a_1^2a_3^2 +a_2^2a_3^2) -a_1^2a_2^2a_3^2\Big]}{\Big[3r_h^8 +2r_h^6\Big(l^2 +\sum_{i=1}^3a_i^2\Big) +r_h^4\Big(l^2\sum_{i=1}^3a_i^2 +a_1^2a_2^2 +a_1^2a_3^2 +a_2^2a_3^2\Big) -l^2a_1^2a_2^2a_3^2\Big]\big[2(\Xi_1\Xi_2 +\Xi_1\Xi_3 +\Xi_2\Xi_3) -\Xi_1\Xi_2\Xi_3\big]}
\ee
as the zero point of the vector field $\phi$.

\begin{figure}[t]
\centering
\includegraphics[width=0.35\textwidth]{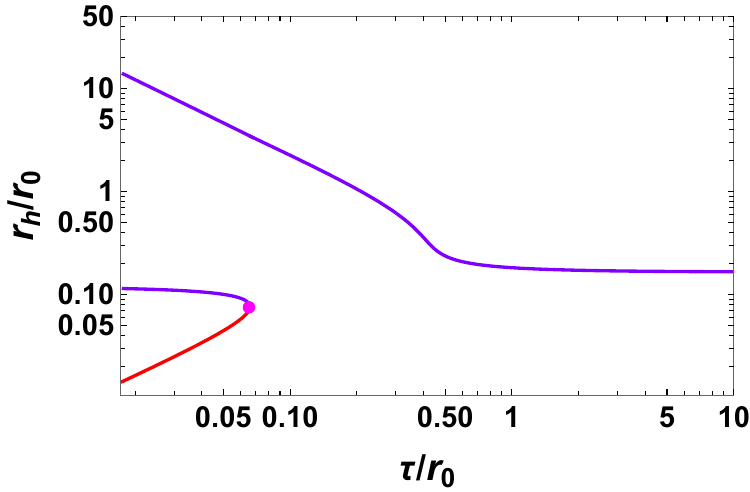}
\caption{Plot of the vector field $\phi^{r_h}$ zero points in the $r_h$-$\tau$ plane for the seven-dimensional triple-rotating Kerr-AdS black hole. The thermodynamically unstable and stable branches, identified by their winding numbers $w = -1$ (red curve) and $w = +1$ (purple curve) respectively, meet at the AP (pink dot). The total topological number is $W = +1$, obtained from the sum of two stable and one unstable state. The chosen parameters are $a_1/r_0 = 0.5$, $a_2/r_0 = 0.25$, $a_3 = 0.2$, and $l/r_0 = 1/3$.
\label{fig4}}
\end{figure}

To investigate the topological structure, we explore the zero points of the vector $\phi^{r_h}$ in the $r_h$-$\tau$ plane for the seven-dimensional Kerr-AdS black hole with a generic configuration of three non-equal rotation parameters, chosen to be $a_1/r_0 = 0.5$, $a_2/r_0 = 0.25$, and $a_3/r_0 = 0.2$, with the AdS radius set by $l/r_0 = 1/3$. The result is presented in Fig. \ref{fig4}. The more symmetric cases, such as $a_1 = a_2 = a_3$ or $a_1 = a_2 \neq a_3$, are not presented here because they yield the same qualitative topological classification and thus provide no new insight.

Upon comparing the topological structures in Fig. \ref{fig4} and Fig. \ref{fig2}, we find that the seven-dimensional triple-rotating and five-dimensional doubly-rotating Kerr-AdS black holes share the same qualitative thermodynamic topology. This universality is manifest in three key aspects: the value of the topological number, the systematic ordering of the zero points, and their asymptotic thermodynamic behavior in the $r_h$-$\tau$ plane.

In summary, the seven-dimensional triple-rotating Kerr-AdS black hole also belongs to the $\widetilde{W}^{1+}$ subclass.

%%%%%%%%%%%%%%%%%%%%%%%%%%%%%%%%%%%%%%%%%%%%%%%%%%%%%%%%%%%%%%%%%%%%%%%%%%%%%%%%%%%%%%%%%%
%%%%%%%%%%%%%%%%%%%%%%%%%%%%%%%%%%%%%%%%%%%%%%%%%%%%%%%%%%%%%%%%%%%%%%%%%%%%%%%%%%%%%%%%%%

%%%%%%%%%%%%%%%%%%%%%%%%%%%%%%%%%%%%%%%%%%%%%%%%%%%%%%%%%%%%%%%%%%%%%%%%%%%%%
\section{Nine-dimensional four-rotating Kerr-AdS black hole case}\label{AppB}
%%%%%%%%%%%%%%%%%%%%%%%%%%%%%%%%%%%%%%%%%%%%%%%%%%%%%%%%%%%%%%%%%%%%%%%%%%%%%

For the thermodynamic topological classification of the nine-dimensional, four-rotating Kerr-AdS black hole, the function $\mathcal{A}_7$ in the generalized off-shell Helmholtz free energy Eq. (\ref{FE}) is given by $\pi^4/3$, which reduces the expression to
\be
\mathcal{F} = \frac{\pi^3(r_h^2 +l^2)[2(\Xi_1\Xi_2\Xi_3 +\Xi_1\Xi_2\Xi_4 +\Xi_1\Xi_3\Xi_4 +\Xi_2\Xi_3\Xi_4) -\Xi_1\Xi_2\Xi_3\Xi_4]}{48r_h^2l^2}\prod_{i=1}^4\frac{r_h^2 +a_i^2}{\Xi_i^2} -\frac{\pi^4}{12r_h\tau}\prod_{i=1}^4\frac{r_h^2 +a_i^2}{\Xi_i} \, .
\ee
Then the components of the vector $\phi$ can be computed as
\bea
\phi^{r_h} &=& \frac{\pi^3[2(\Xi_1\Xi_2\Xi_3 +\Xi_1\Xi_2\Xi_4 +\Xi_1\Xi_3\Xi_4 +\Xi_2\Xi_3\Xi_4) -\Xi_1\Xi_2\Xi_3\Xi_4]}{24r_h^3l^2\prod_{i=1}^4\Xi_i^2}\Bigg\{4r_h^{10} +3r_h^8\left(l^2 +\sum_{i=1}^4a_i^2\right) +2r_h^6\Bigg[l^2\sum_{i=1}^4a_i^2 +\Big(a_1^2a_2^2  \nn \\
&&+a_1^2a_3^2 +a_1^2a_4^2 +a_2^2a_3^2 +a_2^2a_4^2 +a_3^2a_4^2 \Big)\Bigg] +r_h^4\Big[l^2\Big(a_1^2a_2^2 +a_1^2a_3^2 +a_1^2a_4^2 +a_2^2a_3^2 +a_2^2a_4^2 +a_3^2a_4^2\Big) +a_1^2a_2^2a_3^2 +a_1^2a_2^2a_4^2  \nn \\
&&+a_1^2a_3^2a_4^2 +a_2^2a_3^2a_4^2 \Big] -l^2\prod_{i=1}^4a_i^2\Bigg\} -\frac{7r_h^8 +5r_h^6\sum_{i=1}^4a_i^2 +3r_h^4(a_1^2a_2^2 +a_1^2a_3^2 +a_1^2a_4^2 +a_2^2a_3^2 +a_2^2a_4^2 +a_3^2a_4^2)}{12r_h^2\tau\Xi_1\Xi_2\Xi_3\Xi_4} \nn \\
&&-\frac{r_h^2(a_1^2a_2^2a_3^2 +a_1^2a_2^2a_4^2 +a_1^2a_3^2a_4^2 +a_2^2a_3^2a_4^2) -\prod_{i=1}^4a_i^2}{12r_h^2\tau\Xi_1\Xi_2\Xi_3\Xi_4} \, , \\
\phi^{\Theta} &=& -\cot\Theta\csc\Theta \, .
\eea
Therefore, the zero point of the vector can be easily given as
\bea
\tau = \frac{2\pi r_hl^2\Xi_1\Xi_2\Xi_3\Xi_4\Big(7r_h^8 +5r_h^6\sum_{i=1}^4a_i^2 +3r_h^4Y +r_h^2Z -\prod_{i=1}^4a_i^2\Big)}{\Big[4r_h^{10} +3r_h^8\left(l^2 +\sum_{i=1}^4a_i^2\right) +2r_h^6\Big(l^2\sum_{i=1}^4a_i^2 +Y\Big) +r_h^4\Big(l^2Y
+Z \Big) -l^2\prod_{i=1}^4a_i^2\Big](2X -\Xi_1\Xi_2\Xi_3\Xi_4)} \, ,
\eea
where
\bea
&&X = \Xi_1\Xi_2\Xi_3 +\Xi_1\Xi_2\Xi_4 +\Xi_1\Xi_3\Xi_4 +\Xi_2\Xi_3\Xi_4 \, , \nn \\
&&Y = a_1^2a_2^2 +a_1^2a_3^2 +a_1^2a_4^2 +a_2^2a_3^2 +a_2^2a_4^2 +a_3^2a_4^2 \, , \nn \\
&&Z = a_1^2a_2^2a_3^2 +a_1^2a_2^2a_4^2 +a_1^2a_3^2a_4^2 +a_2^2a_3^2a_4^2 \, . \nn
\eea

%%%%%%%%%%%%%%%%%%%%%%%%%%%%%%%%%%%%%%%%%%%%%%%%%%%%%%%%%%%%%%%%%%%%%%%%%%%%%%%%%%%%%%%%%%
\begin{figure}[t]
\centering
\includegraphics[width=0.35\textwidth]{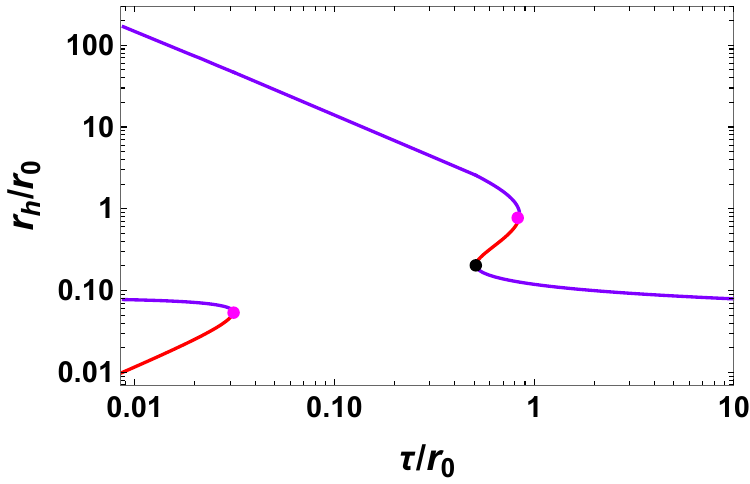}
\caption{Zero points of the vector field $\phi^{r_h}$ in the $r_h-\tau$ plane for the nine-dimensional, four-rotating Kerr-AdS black hole. The thermodynamically unstable and stable branches, characterized by their respective winding numbers $w = -1$ (red curve) and $w = 1$ (purple curve), converge at the AP (pink dot) and the GP (black dot). The global topological number is $W = 1$, which is the sum of contributions from two stable (+1) and one unstable (-1) states. Parameters are fixed as $a_1/r_0 = 1/3$, $a_2/r_0 = 0.25$, $a_3/r_0 = 0.2$, $a_3/r_0 = 1/6$ and $l/r_0 = 1$.
\label{fig5}}
\end{figure}
%%%%%%%%%%%%%%%%%%%%%%%%%%%%%%%%%%%%%%%%%%%%%%%%%%%%%%%%%%%%%%%%%%%%%%%%%%%%%%%%%%%%%%%%%%

To probe the topological structure, we map the zero points of the vector field $\phi^{r_h}$ in the $r_h$-$\tau$ plane for a nine-dimensional Kerr-AdS black hole. A generic case with non-equal parameters ($a_1/r_0 = 1/3$, $a_2/r_0 = 0.25$, $a_3/r_0 = 0.2$, $a_4/r_0 = 1/6$, and $l/r_0 = 1$) is examined, with the result depicted in Fig. \ref{fig5}. The more symmetric configurations, such as the fully equal or partially equal cases, are not shown as they fall into the same topological class and do not alter our qualitative conclusions.

Remarkably, the topological structures of the nine-dimensional (four-rotating) and five-dimensional (doubly-rotating) Kerr-AdS black holes, compared in Fig. \ref{fig5} and Fig. \ref{fig2}, are qualitatively identical. This topological universality is evident from their shared topological number, the common sequence of zero points, and their parallel asymptotic thermodynamic behavior in the $r_h$-$\tau$ plane.

Consequently, this analysis confirms that the nine-dimensional, four-rotating Kerr-AdS black hole resides in the $\widetilde{W}^{1+}$ topological subclass. This result aligns with the previously established five-dimensional, doubly-rotating case (in the main text) and is consistent with the seven-dimensional, triple-rotating black hole detailed in Appendix \ref{AppA}.

\twocolumngrid

\end{document}